# Artificial Intelligence Reshapes Microwave Photonics

*Peng Li,[1,2] Xihua Zou,[1,2,*] Jia Ye,[1,2] Wei Pan,[1,2] and Lianshan Yan[1,2]*

## Affiliations:

[1]Center for Information Photonics and Communications, School of Information Science and Technology, Southwest Jiaotong University, Chengdu 611756, China
[2]Key Laboratory of Photonic-Electronic Integration and Communication-Sensing Convergence, Southwest Jiaotong University, Chengdu 611756, China
*Corresponding author: zouxihua@swjtu.edu.cn

## Abstract:

As a rapidly emerging interdisciplinary field that intrinsically integrates microwave and photonics, microwave photonics (MWP) provides disruptive solutions to overcome the fundamental bandwidth of conventional electronic systems. By exploiting the inherently ultra-wide bandwidth and low-loss characteristics of photonic technologies, MWP enables the generation, transmission, processing, and detection of microwave, millimeter-wave, and terahertz signals. Representative breakthroughs include fully photonic microwave radar systems, photonic analog-to-digital converters with bandwidth up to 320 GHz, and photonic wireless communication systems achieving data rate as high as 616 Gbit/s. Meanwhile, the rapid growth of artificial intelligence (AI) is reshaping scientific research, engineering, and daily life in unprecedented ways, such as AI for science/engineering and AI co-scientist/assistant. Correspondingly, AI is profoundly reshaping MWP in all aspects, ranging from signal generation, transmission to signal processing and detection. AI has revolutionized the design, simulation, fabrication, testing, deployment, and maintenance of MWP systems, delivering autonomous operation and exceptional efficiency beyond traditional systems. Motivated by these developments, this Review Paper provides the first comprehensive overview of AI-enabled MWP, systematically summarizing the state-of-the-art advances and presenting insights for both the academic community and the broader public.

## 1. Introduction

Microwave photonics (MWP) is an interdisciplinary field that integrates microwave and photonic techniques to generate, transmit, process, and detect microwave, millimeter-wave, and terahertz-wave signals [1-5]. Owing to the intrinsic advantages (e.g., low insertion loss, large instantaneous bandwidth, and immunity to electromagnetic interference) of photonics, MWP has found numerous applications such as broadband wireless communication [6-9], ultra-wideband radar [10-14], high-resolution ranging and sensing [15-18], and modern electronic warfare [19-21].

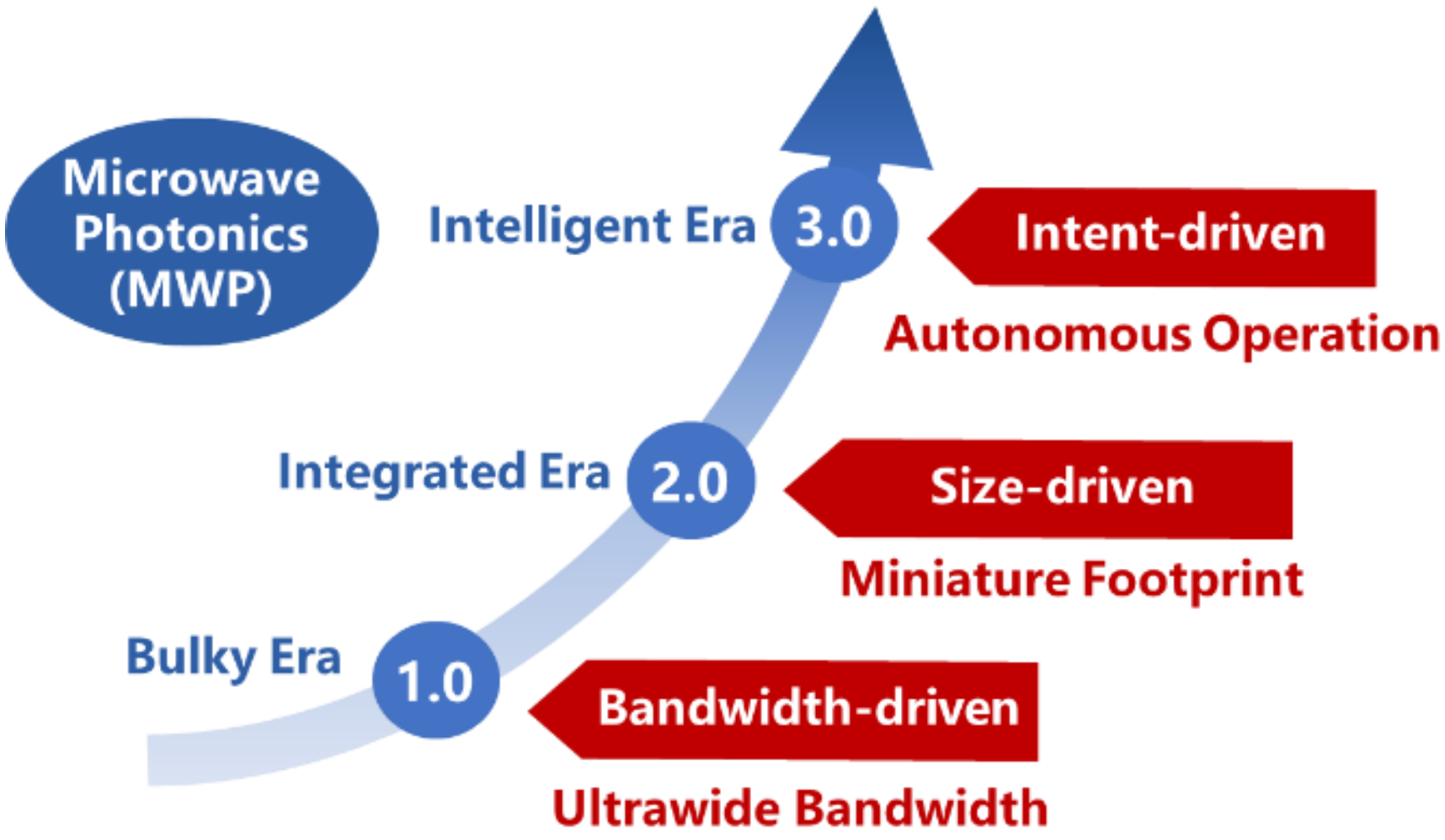

**Figure 1.** Evolution roadmap of microwave photonics (MWP)

Originated from 1970s [1], [22-24], MWP has undergone an explicit evolution through three distinct eras—**MWP 1.0 (bulky era)**, **MWP 2.0 (integrated era)**, and the current **MWP 3.0 (intelligent era)**—as illustrated in **Fig. 1**. This evolution roadmap highlights several technological milestones, including the ultra-wide bandwidth of photonics, the miniaturized footprint of photonic integrated circuits (PICs), and the autonomous operation of artificial intelligence (AI). The MWP 1.0 era is a bandwidth-driven phase, which utilizes of discrete optical or photonic components in benchtop architectures to achieve ultrawide bandwidth, but at the cost of bulky size, high power consumption, and limited scalability. In the size-driven MWP 2.0 era, photonic integration technologies—such as silicon photonics [25-27], indium phosphide (InP) [4], [28-29], and silicon nitride [30-31] platforms— have enabled the miniature footprint of MWP subsystems or entire systems onto a single chip, thereby enhancing stability, integration density, and system reliability. However, as integration density and functional complexity continue to grow, integrated MWP systems increasingly encounter efficiency and optimization bottlenecks. The large-scale monolithic integration of lasers, modulators, filters, and detectors introduces intricate interdependencies among multiple parameters and objectives, making adaptive and real-time optimization highly challenging in complex or dynamically varying environments.

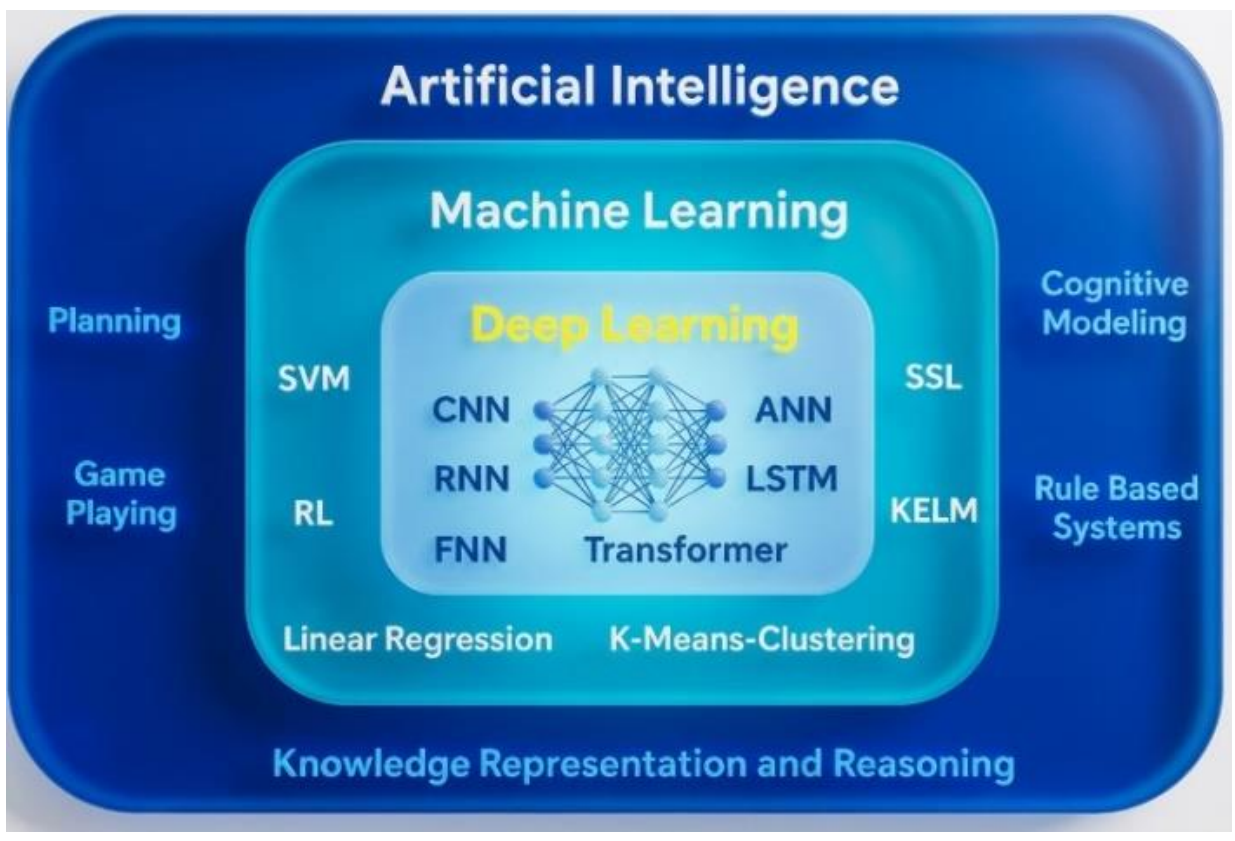

**Figure 2.** Hierarchy of artificial intelligence (AI), machine learning (ML), and deep learning (DL).

The rise of AI provides a transformative framework to address these challenges by learning high-order implicit relationships between inputs and outputs directly from data training, thus overcoming the constraints imposed by limited analytical models, incomplete prior knowledge, and scarce experimental or computational resources. The hierarchical relationship among AI, machine learning (ML), and deep learning (DL) is illustrated in **Fig. 2**. ML is a core subset of AI and is generally classified into supervised learning (learning from labeled data), unsupervised learning (finding structure in unlabeled data), and reinforcement learning (RL) where agents learn optimal strategies through trial and reward. Classical ML approaches include support vector machines (SVM), k-nearest neighbors (KNN), self-supervised learning (SSL), and K-means clustering, and so on. Deep learning (DL) is an advanced branch of ML that uses artificial neural networks (ANNs) with multiple layers to learn hierarchical features. Typical DL architectures involve feedforward neural networks (FNNs), convolutional neural networks (CNNs) and deep residual networks (DRNs), generative adversarial networks (GANs), recurrent neural networks (RNNs), long short-term memory (LSTM) networks, and Transformer. Within this framework, ML develops a class of algorithms that automatically extract patterns and predictive relationships from experimental or simulated data [31-33], while DL utilizes multi-layer neural networks to capture highly nonlinear relationships and to abstract complex physical behaviors [34-35].

Recent comprehensive reviews [36–39] have highlighted how AI has revolutionized photonics and related fields by accelerating forward modeling, inverse design, and real-time optimization at both device and system levels. Extending these capabilities to MWP, AI introduces transformative advantages across a wide range of approaches and applications, such as nonlinear distortion compensation for analog photonic links [40-41], adaptive beamforming for photonic phased arrays [42-43], and data-driven calibration for PICs [44-45]. Consequently, we stepped into the intent-driven MWP 3.0, characterized by the convergence of AI's autonomous operation, PIC's miniaturization, and photonics' ultra-large bandwidth. Beyond MWP 2.0 emphasizing hardware miniaturization and integration, MWP 3.0 further incorporates data-driven modeling, adaptive control, and dynamic optimization for microwave photonic chips and systems. This review aims to provide a comprehensive perspective on how AI reshapes MWP. As illustrated in **Fig. 3**, it is organized into three major sections—AI-enabled microwave signal generation, transmission, and processing and detection. Finally, the remaining challenges and future research perspectives toward fully autonomous and intelligent MWP platforms are discussed.

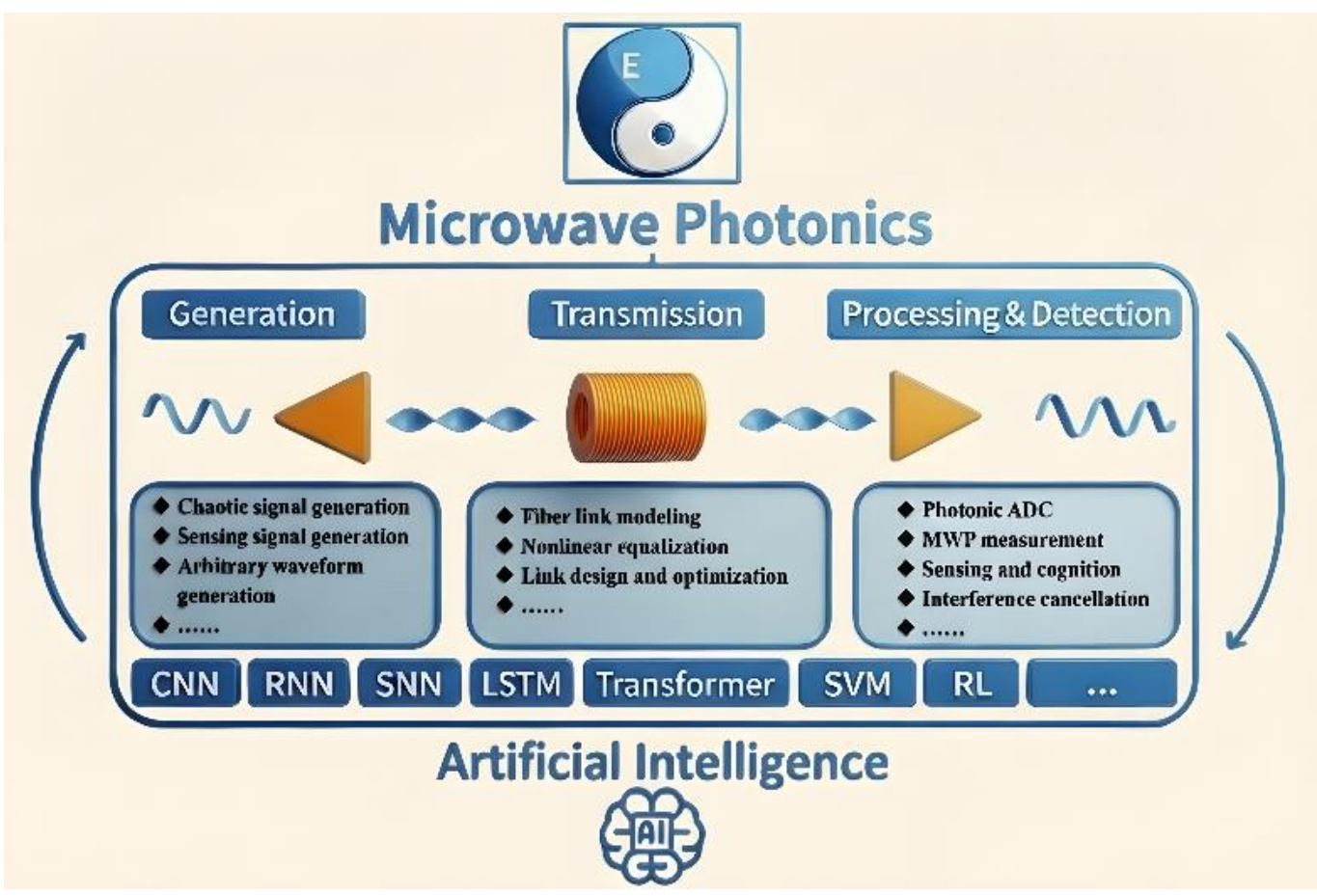

**Figure 3**. Architecture of "AI reshapes MWP".

## 2. AI reshapes MWP systems

### 2.1. AI for MWP signal generation

Conventional MWP signal generators—such as optoelectronic oscillators (OEOs) [46-50], optical heterodyning approaches [51-53], and frequency-to-time mapping systems [54-55]—are capable of generating high-frequency or wideband signals with low phase noise. However, OEOs often require precise manual calibration and active stabilization, as their oscillation states are highly sensitive to environmental fluctuations and initial conditions. The emergence of AI has introduced for a data-driven and self-adaptive paradigm for MWP signal generation [56-59]. In these frameworks, AI not only enables real-time and flexible manipulation of device and system, but also reveals the complex nonlinear mapping relationships between optical and electrical domains, which empowers intelligent generation of sophisticated and reconfigurable signals or waveforms for communication, radar, and sensing applications.

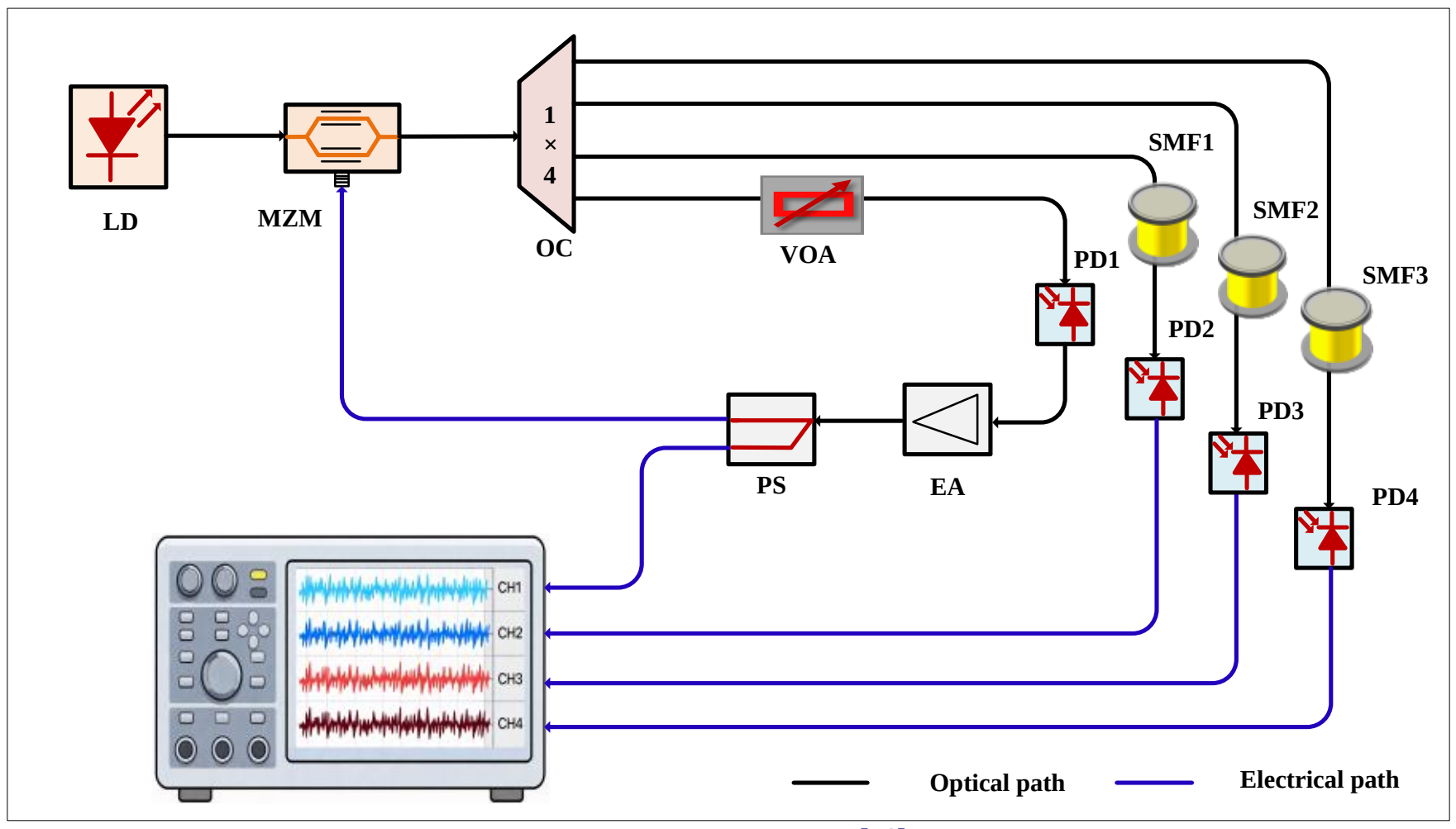

**Figure 4**. Schematic diagram of the NOEO-based photonic accelerator [56]. (LD: laser diode; MZM: Mach-Zehnder modulator; OC: optical coupler; VOA: variable optical attenuator; SMF: single mode fiber; PD: photodetector; EA: electrical amplifier; PS: power splitter).

An AI-enabled **chaotic signal** generation approach has been proposed based on a nonlinear optoelectronic oscillator (NOEO) [56], as shown in Fig. 4. The NOEO, described by a set of coupled nonlinear delay differential equations, leverages feedback-induced chaos generated in a cavity incorporating a Mach–Zehnder modulator (MZM). By precisely balancing the optical gain and the cavity nonlinearity, the NOEO generates four parallel and mutually orthogonal chaotic sequences with a bandwidth of 18.18 GHz and a permutation entropy of 0.9983, indicating near-maximal randomness and complexity. These ultra-broadband chaotic signals are the fundamental carriers of an all-photonic reinforcement-learning (RL) accelerator capable of tackling representative decision-making tasks such as the multi-armed bandit (MAB) and Tic-Tac-Toe (TTT) problems. Experimentally, the system exhibits uncorrelated outputs for time delays exceeding 100 ns, providing orthogonal chaotic channels through simple fiber-delay diversity.

Another AI-enabled solution for **sensing signal** generation is developed through machine-learning (ML)-assisted control of period-one (P1) microwave oscillations in semiconductor lasers with external optical feedback (EOF) [57]. Traditionally, the oscillation frequency is determined by solving the Lang–Kobayashi (L–K) rate equations, which describe the coupled evolution of electric field, carrier density, and feedback delay. However, iterative numerical solutions of these delay differential equations are computationally intensive, making real-time parameter tuning impractical. This limitation can be effectively overcome by training a feedforward neural network (FNN) model to emulate the L–K model. The FNN-L–K solver takes the system control parameters (SCPs)—injection current, cavity length, and feedback strength—as inputs, and then predicts the corresponding P1 frequency with high fidelity. Using a dataset of 50,000 simulated samples, a 2×20 neuron network has achieved a mean squared error below $10^{-5}$, corresponding to sub-megahertz prediction accuracy. When being combined with a gradient-descent optimization algorithm, the model rapidly determines optimal SCPs for a target frequency within milliseconds on a CPU or tens of nanoseconds on an FPGA, representing an efficiency improvement of more than three orders of magnitude compared with direct numerical solvers.

Next, an AI-enabled programmable **optical frequency comb (OFC)** generation approach has been proposed based on nonlinear spectral broadening in a highly nonlinear fiber [58]. The system utilizes an electro-optic OFC as the seed source, followed by neural-network-controlled spectral phase modulation and nonlinear propagation to achieve flexible comb shaping. By using a 1D-ResNet architecture, the inverse mapping between the target comb spectrum and the required seed temporal profile can be efficiently learned, enabling accurate modeling of high-order nonlinear effects such as self-phase modulation and dispersion. Experimentally, the system demonstrates the programmable generation of diverse spectral envelopes, including Gaussian, parabolic, Cauchy, Laplace, and Gaussian mixture profiles with an error of $10^{-1}$–$10^{-2}$.

## 2.2. AI for MWP transmission

Traditional MWP transmission links mostly rely on deterministic physical models, fixed digital signal processing (DSP) algorithms, and independently optimized link components, making them increasingly inadequate in handing nonlinear distortions, diverse environmental conditions, and the rising complexity of fiber–wireless networks. Recent advances in AI have introduced a data-driven paradigm for intelligent MWP transmission [60-77]. By learning signal evolution dynamics directly from data, AI enables efficient nonlinear impairment compensation and joint optimization of transmitters and receivers in real-time. Consequently, AI-enabled MWP transmission systems achieve enhanced accuracy, autonomous adaptability, and scalable capacity.

### 2.2.1 Fiber link modeling

AI has emerged as a powerful tool for modeling, optimizing, and predicting the behavior of MWP or radio-over-fiber (RoF) links. Traditional link modeling methods—such as the split-step Fourier method (SSFM)—accurately describe optical signal transmission through fibers by dividing the link into many small segments and solving them sequentially [59]. However, this approach incurs high computational complexity and limited flexibility when adapting to different signal formats, data rates, or transmission distances. In contrast, AI-enabled modeling, particularly using DL, provides a new and powerful way to model fiber links. Instead of relying solely on explicit physical equations, DL learns the transmission behavior in the fiber directly from data, enabling faster simulation an improved adaptability to complex nonlinear effects [60-67].

A data-driven **fiber link model** based on a deep neural network (DNN) with a multi-head attention mechanism has been proposed to accurately and efficiently predict the propagation evolution of optical signals along fiber link [63]. The multi-head attention mechanism enables the parallel extraction of diverse signal features, thereby effectively modeling long-range dependencies and complex distortions in time-series data. This approach represents a significant shift from the sequential processing paradigm employed by BiLSTM architectures It performs well across various data rates, modulation formats, and transmission distances, showing strong generalization capability. By learning both explicit and implicit relationships between input and output signals under diverse conditions, the model can precisely predict transmission characteristics, like a 16-QAM signal at 160 Gb/s transmitted over100 km of fiber.

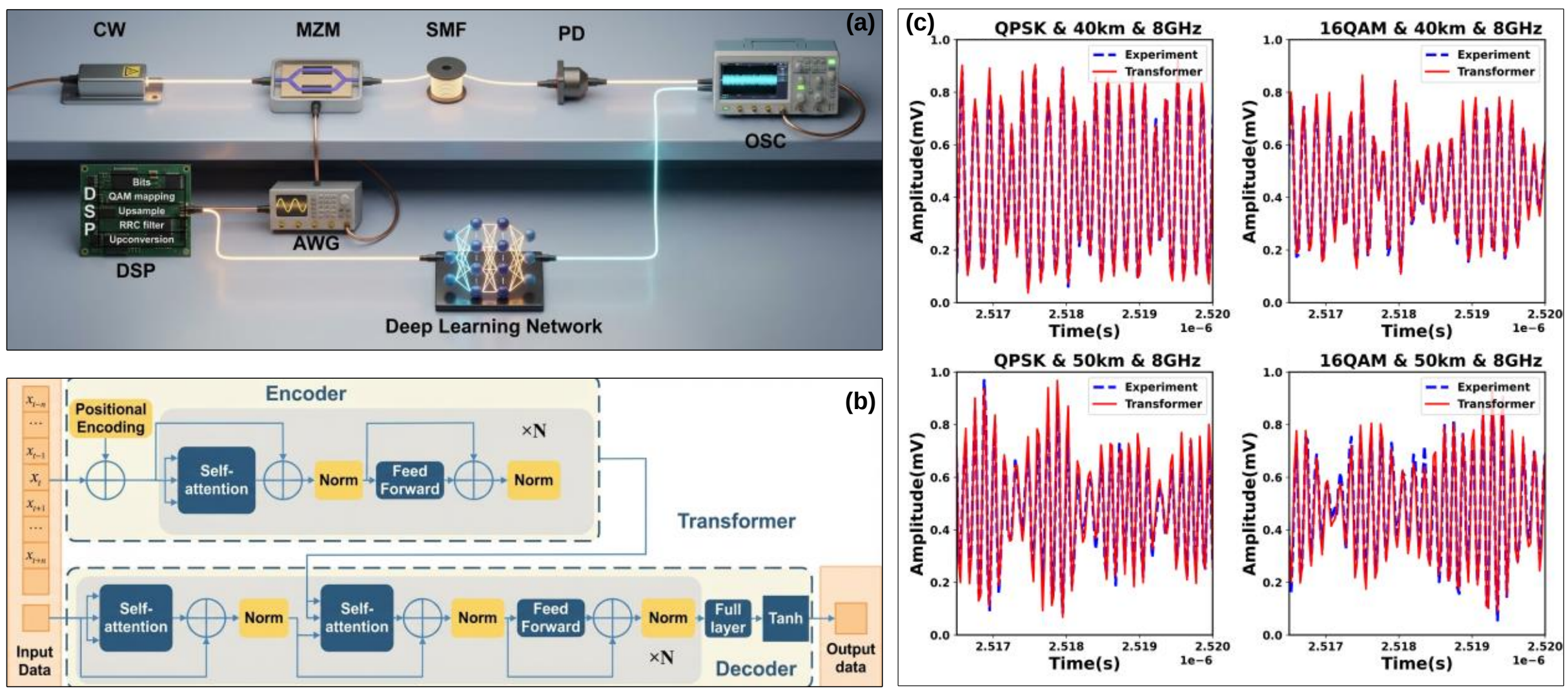

**Figure 5**. (a) Transformer-based RoF transmission link, (b) improved Transformer architecture for RoF link modeling, and (c) comparison of output waveform amplitudes obtained from experimental measurements and transformer-based predictions [67]. (RRC: root raised cosine; AWG: arbitrary waveform generator; CW: continuous wave; PC: polarization controller; MZM: Mach-Zehnder modulator; SMF: single mode fiber; PD: photodetector; OSC: Oscilloscope).

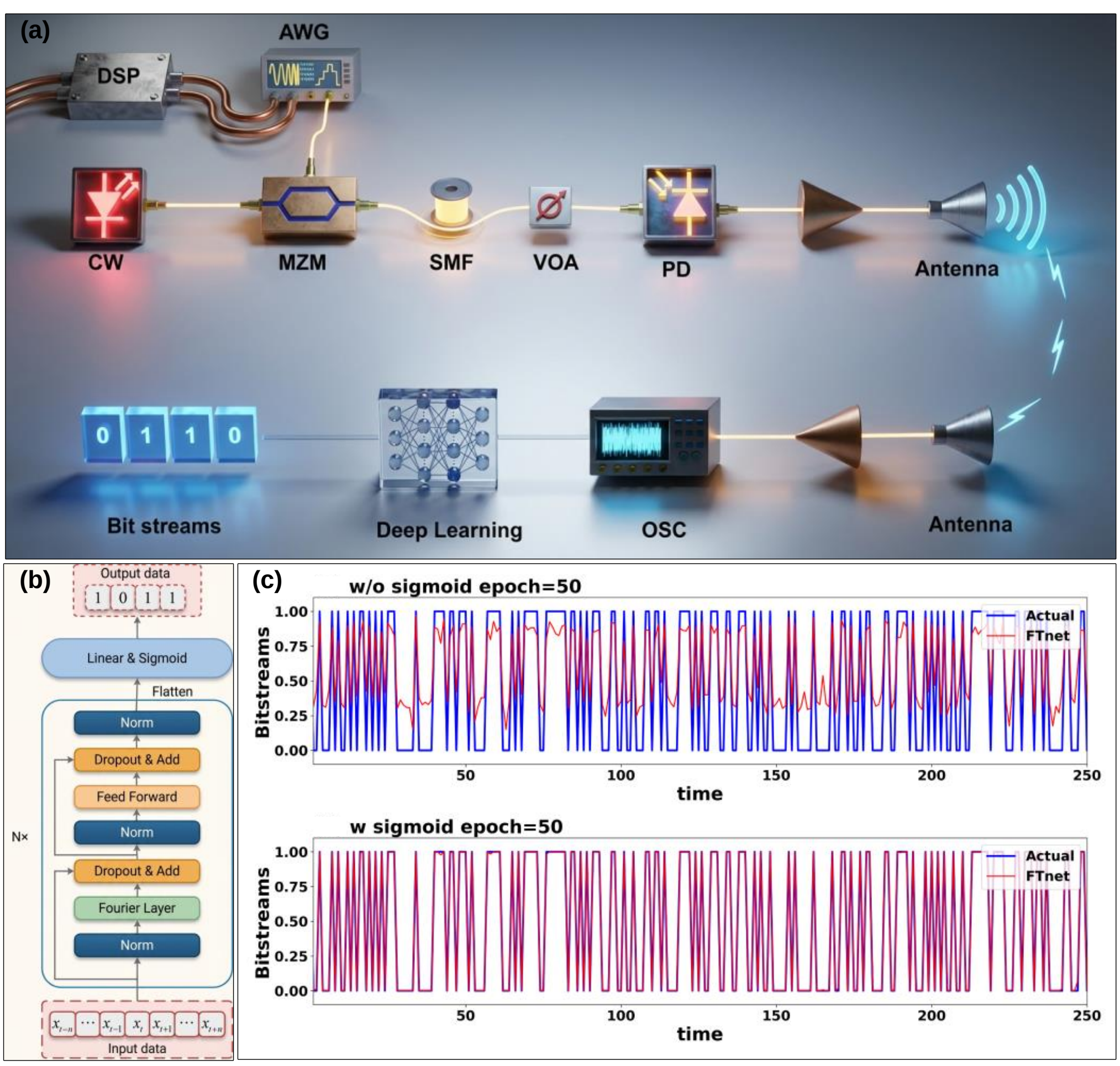

**Figure 6**. (a) RoF system based on a based on a Fourier Layer Transformer network (FTnet), (b) architecture of the Ftnet model with N encoder sublayers, and (c) experimental results after 50 training epoches with and without the sigmoid function [72]. (DSP: digital signal processing; AWG: arbitrary waveform generator; CW: continuous wave; MZM: Mach-Zehnder modulator; SMF: single mode fiber; VOA: variable optical attenuator; PD: photodetector; OSC: Oscilloscope).

Another **Transformer-based link model** has also been developed [67] for RoF links, as shown in **Fig. 5**. Unlike previous solutions trained only with simulated data, this approach uses experimentally measured data from a real RoF link for training. It learns the impacts of different physical effects (e.g., attenuation, dispersion, noise, and nonlinearity) on signal transmission. For an RoF link deploying a 2-GBaud 16QAM signal at 8 GHz carrier frequency, the Transformer model has achieved an extremely low normalized mean square error (NMSE = 0.000819), which is far below the acceptable threshold of 0.02. Compared with other AI models such as fully connected (FCNN), convolutional (CNN), and generative adversarial networks (GANs), the Transformer shows both higher prediction accuracy and faster computational efficiency. More importantly, the model maintains high performance under various conditions, including fiber lengths from 10 to 50 km, microwave carrier frequencies from 4 to 10 GHz, and SNRs from 4 to 20 dB.

### 2.2.2 Nonlinear effect equalization of transmission links

RoF systems face persistent challenges from nonlinear effects, including laser frequency drift, imperfections in modulators, fiber-induced nonlinearity, and photodetector saturation. These effects severely distort transmitted signals and degrade overall system performance. Conventional equalization techniques—such as Volterra equalizers and least mean square (LMS) algorithms [68]—struggle to handle these complex time-varying nonlinear distortions. DL-based methods have emerged as promising alternatives, capable of learning nonlinear mappings directly from experimental or simulated data without requiring explicit analytical models. These data-driven approaches greatly enhance the flexibility and precision of signal equalization, thereby remarkably improving the performance of high-speed RoF communications [69-72].

A **DL-based RoF receiver** using a long short-term memory (LSTM) network has been proposed to overcome phase noise and nonlinear distortions in unlocked heterodyne RoF links [70]. The system, known as Deep Learning Detection with Reference Tone (DLD-RT), jointly utilizes both a reference tone and phase-distorted in-phase/quadrature (IQ) samples as network inputs. This configuration enables the LSTM network to learn the temporal evolution of the signal and the influence of phase noise induced by nonlinear mixing. Simulation results demonstrate a substantial reduction in bit error rate (BER) from $10^{-1}$ to $10^{-5}$, outperforming other DL architectures such as the multilayer perceptron (MLP) and CNN. Compared with traditional self-homodyning receivers, the LSTM-based model exhibited superior robustness to phase noise and enhanced stability under varying frequency spacing conditions. Unlike conventional digital signal processing (DSP) techniques, which often fail to accurately model complex temporal correlations, the DL-based framework can adaptively capture time-varying nonlinear dynamics and significantly enhance signal detection accuracy.

Another improved method— the so-called **data-driven digital demodulator** for RoF systems, has been proposed based on a Fourier Layer Transformer network (FTnet) [72], as shown in **Fig. 6**. This approach takes advantages of DL to directly process and interpret complex optical signals in the time domain without relying on traditional step-by-step DSP. The FTnet model also combines Fourier analysis with a Transformer network, which helps understand how signal behaves in the frequency domain. By bringing the two functions together, the FTnet can accurately learn both time-domain and frequency-domain distortions caused by nonlinear effects, fiber dispersion, and channel interference. Instead of using multiple manually designed correction steps like frequency offset adjustment or equalization, FTnet performs end-to-end learning and thus automatically recovers the original data directly from distorted signals. The 10 GHz RoF link experimentally shows a 3–6 dB improvement in receiver sensitivity and a BER below the 7% FEC limit ($3.8\times10^{-3}$).

### 2.2.3 Transmission link design and optimization

AI is fundamentally transforming the design and optimization of RoF systems particularly in hybrid fiber and wireless transmission links. Traditional design strategies independently optimize each subsystem—such as the transmitter, channel, and receiver [72-73]—which often leads to suboptimal link performance when facing multiple coupled nonlinear distortions and multiple-user or multiple-channel interference. In contrast, AI-enabled end-to-end (E2E) methods treat the entire communication link as a single unified system [74-77]. By leveraging DL, both the transmitter and receiver can be jointly trained to automatically find the globally optimal configurations for the entire link. This data-driven method allows the system to adaptively handle complex links and heterogeneous distortions.

An **E2E system** is designed to use artificial neural networks (ANNs) for the transmitter (T-ANN), receiver (R-ANN), and channel model (ACM) [74]. The link configuration is similar to that depicted in Fig. 6(a). During the training stage, the architecture of the single-user E2E framework is presented in Fig. 7. The ACM acts like a virtual replica of the physical communication channel. It learns to simulate both linear and nonlinear effects in fiber–wireless links, such as inter-symbol interference, dispersion, and optical nonlinearities, using a special dual-branch neural network. A major improvement is the two-step transfer learning strategy for ACM, the first training by measured experimental data and the second re-training by AI-generated signals. This design enables end-to-end backpropagation across the entire system, allowing both the transmitter and the receiver to self-optimize in a fully automated manner. In a millimeter-wave-over-fiber link carrying 46.2 Gbit/s, a 3.5 dB or 1.5 dB improvement on receiver sensitivity is achieved for single-user or three-user case.

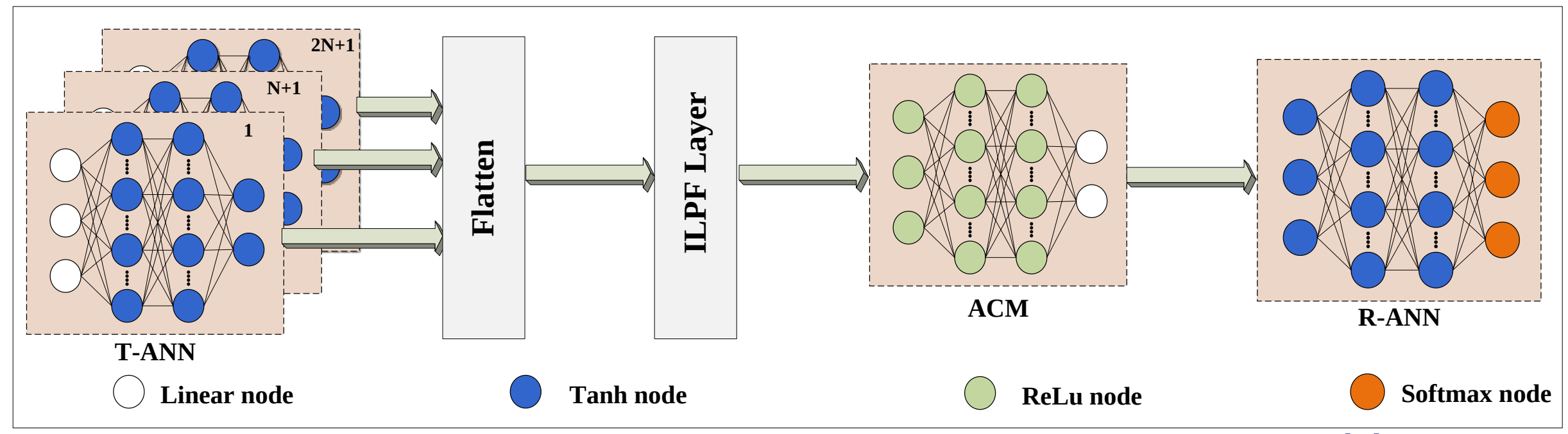

**Figure 7**. Structures of deep-learning-based single-user E2E framework in the training phase [74].

Another **data-driven E2E RoF transmission link** using self-supervised learning (SSL) has also been demonstrated [77]. Rather than relying on traditional DSP, this framework uses four deep neural networks: TransNN for symbol mapping, SamplNN for upsampling, ChannelNN for channel modeling, and ReceivNN for demodulation. The ChannelNN, a differentiable Transformer-based model, enables the joint optimization of the RoF link via gradient-based methods by accurately characterizing fiber nonlinearities, chromatic dispersion, and hardware distortions. Compared with traditional DSP-based methods, a 3.5 dB enhancement in receiver sensitivity along with improved operational stability is observed in simulations for a 25 km RoF link. Moreover, conventional demodulators fail to decode the AI-generated signals, indicating inherent physical-layer security and enhanced adaptability of the proposed system.

## 2.3. AI for MWP signal processing & detection

Conventional MWP signal processing techniques—such as photonic analog-to-digital conversion (ADC) [78-80], filtering [81-83], microwave measurement [84-88], multi-parameter sensing [89-91], and interference cancellation [92-94]—has demonstrated strong capability for processing high-frequency and wideband signals for advanced communication, radar, and sensing systems. However, these methods are increasingly limited by fixed system architectures, complicated calibration procedures, and long-term performance instability. Recently, AI-enabled MWP signal processing has emerged as a transformative paradigm. By leveraging ML or DL, MWP systems can autonomously adjust internal parameters and perform adaptive signal processing and detection. This integration enables intelligent photonic processing of microwave signals featuring larger bandwidth, higher accuracy, faster response, and improved efficiency.

### 2.3.1 Photonic analog-to-digital conversion (ADC)

Photonic analog-to-digital conversion (ADC) offers ultra-high sampling rates and exceptionally wide bandwidth [78-80], while purely electronic ADC suffers from bandwidth limitations and timing jitter. Nevertheless, photonic ADC still faces challenges such as nonlinear distortions and channel mismatches. To overcome these issues, DL-powered photonic ADC (DL-PADC) has been proposed, integrating photonics, electronics, and AI within a unified framework [95-99].

A representative **DL-PADC architecture** is proposed by integrating the photonic front-end, electronic quantization, and DNN correction into a unified system [95]. Besides ultra-wide bandwidth and low timing noise of photonics, the system takes advantages of DL to correct hardware errors. Two cascaded neural networks are employed, the linearization network for eliminating distortions caused by the electro-optic modulator and the matching network for correcting timing errors between multiple sampling channels. A prototype with two channels operating at 20 GSample/s is demonstrated. After applying the DL correction, the effective number of bits (ENOB) is improved from 4.6 to 7.3 bits for input signals up to 21 GHz, and the spurious-free dynamic range (SFDR) is reached 71 dB. By using an ultra-low-jitter mode-locked laser and a high-precision data board, the ENOB is further increased to 9.24 bits at 23 GHz, which is very close to the theoretical limit.

Another example is the application of **AI–PADC** in a photonic radar receiver for automatic target recognition (ATR) [96], as shown in **Fig. 8**. A 10 GHz-bandwidth photonic ADC is firstly used to sample radar echo signals, producing range profiles (RPs) with centimeter-level resolution. The CNN then automatically analyzes these RPs and extracts deep spatial features. In experiments with four small metallic and aerial objects, a recognition accuracy of 93.05% at a range resolution of 7.5 cm is achieved, which is approximately 25% higher than achieved by traditional electronic radar systems. Therefore, the combination of a photonic front-end for high-speed signal capture and an AI-enabled processing for intelligent learning can enable radar system faster and intelligent.

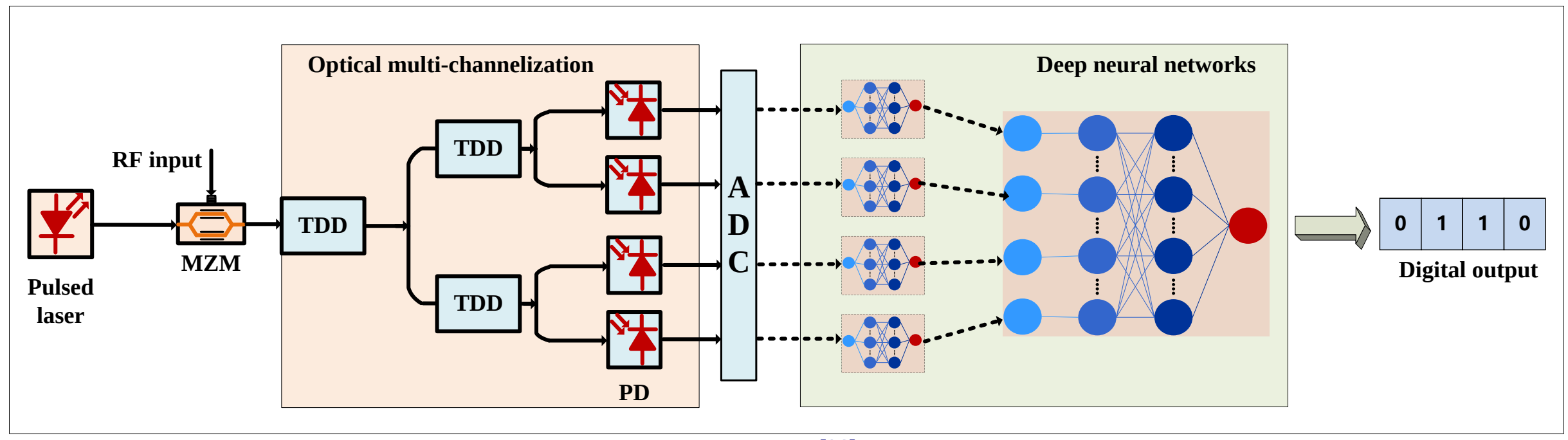

**Figure 8**. Schematic diagram of the DL-PADC architecture [96]. (MZM: Mach-Zehnder modulator; TDD: time-divided demultiplexer; PD: photodetector).

### 2.3.2 Microwave photonics measurement

Photonics for microwave measurement has been widely used in radar, wireless communication, and electronic warfare [84-88]. However, traditional photonic measurement systems still face challenges such as nonlinear system responses, calibration drift, and limited adaptability, which degrade the measurement accuracy and long-term reliability in practical environments. Recently, AI has emerged as a powerful data-driven method to overcome these issues. ML and DL algorithms can be applied to improve both photonic microwave instantaneous frequency measurement (IFM) [100-104] and angle-of-arrival (AOA) estimation [105-107].

A representative example is the **stacking ensemble learning-assisted photonic IFM system** [100]. The system uses an MZM, a phase modulator (PM), and dual photodiodes with adjustable optical and electrical delays. By combining several regression models (e.g., Random Forest, Gradient Boosting, and Support Vector Machines), the ensemble learning method can accurately predict frequency values from amplitude comparison functions (ACFs). Experimental results show that a wide frequency measurement ranges from 1 to 40 GHz with an average error below 5 MHz, while maintaining robust performance even when the input signal power is as low as -30 dBm.

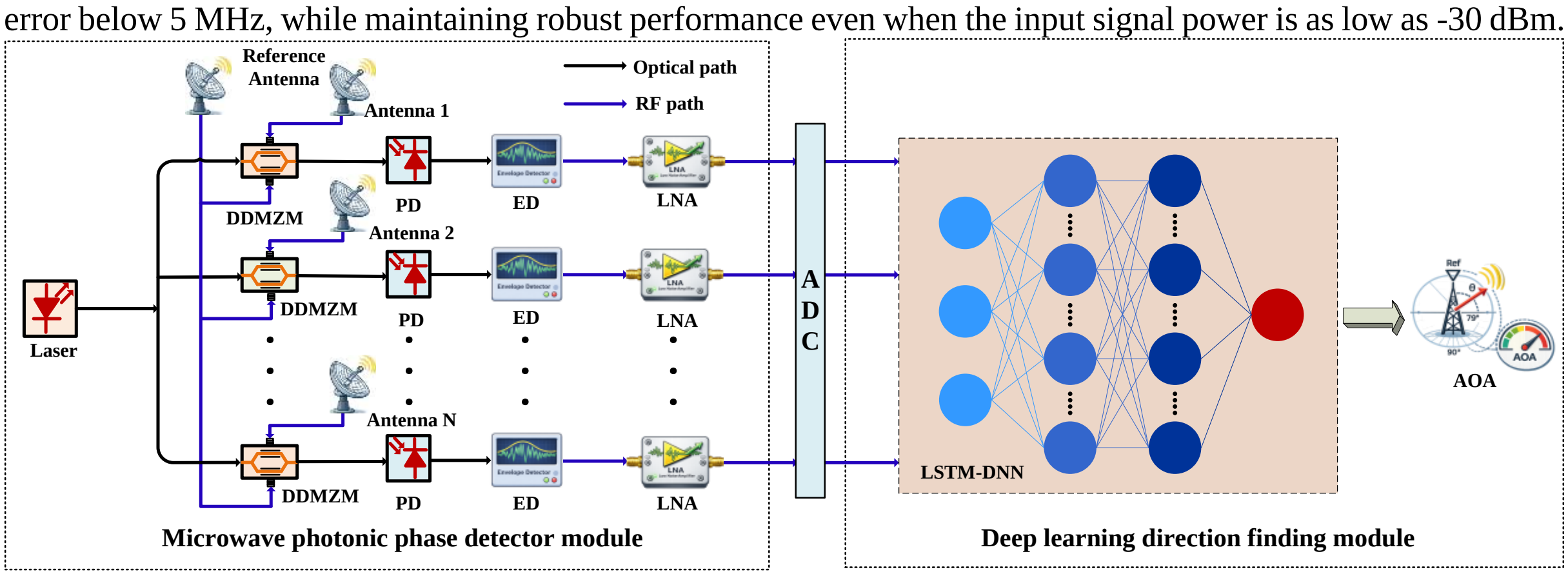

**Figure 9**. Schematic diagram of the DL-assisted AOA estimation system [105]. (DDMZM: dual-drive Mach Zehnder modulator; PD: photodetector; ED: envelope detector; LNA: low noise amplifier).

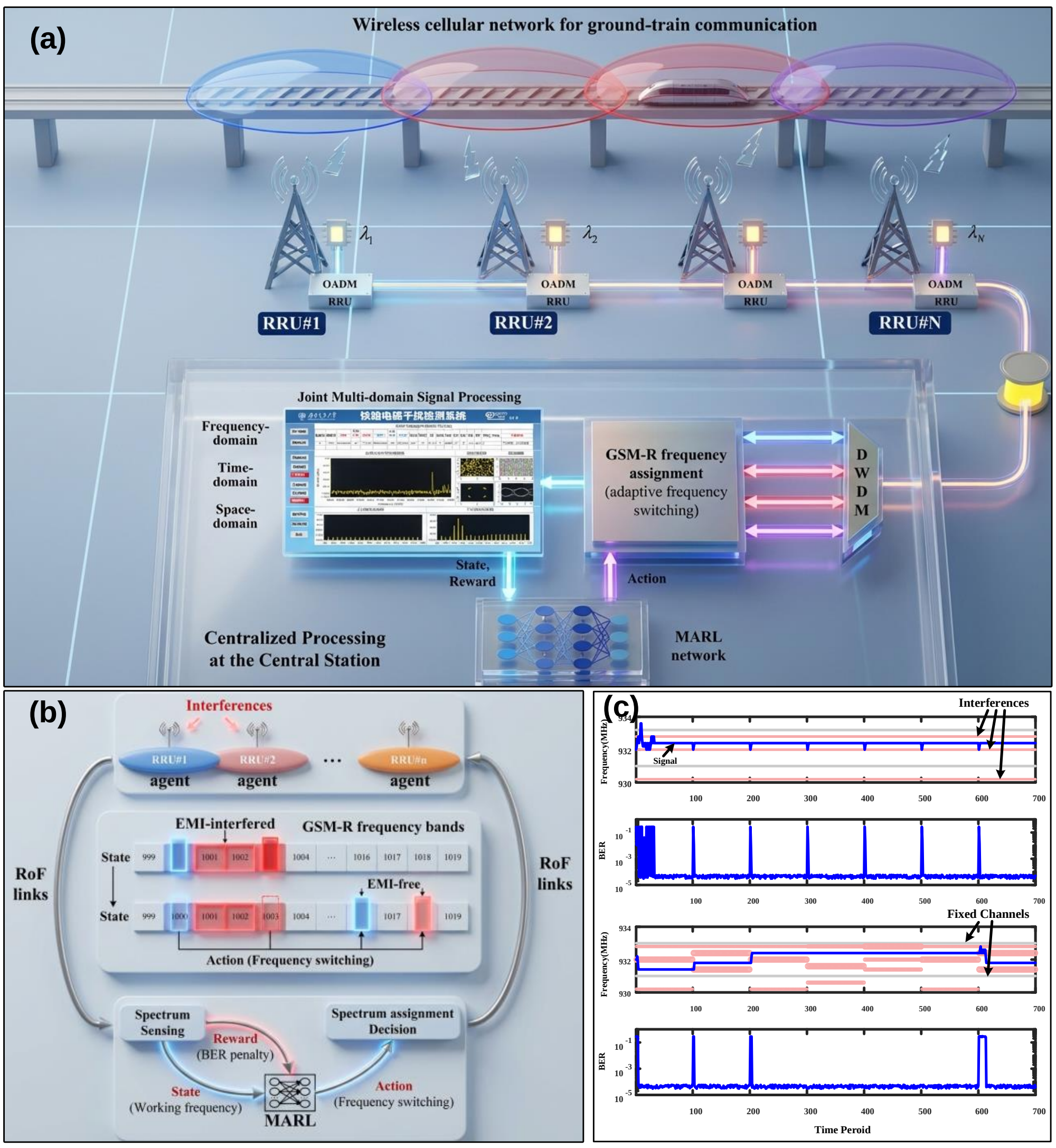

**Figure `10**. (a) Intelligent electromagnetic sensing system enabled by reinforcement learning, (b) schematic illustration of real-time and self-adptive EMI avoidance, and (c) actions executed and corresponding demodulated BERs over time periods under static periodic interference and dynamic abrupt interference scenarios [108]. (OADM: optical add-drop multiplexer; RRU: remote radio unit; DWDM: dense wavelength division multiplexing; MARL: multi-agent reinforcement learning).

Another notable example is the **DL–assisted long-baseline photonic AOA estimation system,** which uses a non-uniform antenna array and dual-drive Mach–Zehnder modulators [105], as shown in **Fig. 9**. In this system, a long short-term memory deep neural network (LSTM-DNN) learns the relationship between the detected voltage signals and the actual signal angle. This allows the system to automatically correct nonlinear effects and eliminates the need for complex phase calibration procedures. The experiments show a very high accuracy with a mean absolute error (MAE) of only 0.1438° across a wide range from −80° to +80°. The system also performs better than traditional DNN and ResNet models requiring only 1.88 ms per measurement.

### 2.3.3 Multi-dimension microwave sensing and cognition

Photonic microwave sensing has emerged as a key technology for real-time, multi-parameter sensing in the fields of communications, biomedicine, and structural monitoring [89-91]. However, traditional MWP sensing methods—such as optical spectrum analysis and curve fitting—are often computationally intensive and time-consuming, limiting their applicability in dynamic scenarios. Recently, ML and DL have brought major improvements to photonic microwave sensing. By learning complex relationships between sensor signals and physical parameters directly from data, AI-enabled approaches enable photonic sensing systems to become faster, more accurate, and more adaptable.

A compelling example is an **intelligent electromagnetic sensing** system for high-speed railway communication, which employs a RoF distributed antenna architecture enhanced by RL [108], as shown in **Fig. 10**. Traditional methods for handling electromagnetic interference (EMI) are usually limited in spatial coverage, operating bandwidth and response speed. Fortunately, the system uses RoF links to collect wideband signals in real time along high-speed railway, and applies a multi-agent RL algorithm (WoLF-PHC). This AI-enabled approach automatically identifies clean frequency channels and coordinates frequency switching among multiple remote radio units (RRUs) along the track, when the communication environment is polluted or attached by EMI. In experiments, the system can successfully avoid stable, dynamic EMIs for high-speed railways (e.g., Chengdu-Chongqing line) in China. Overall, this system exhibits significantly enhanced reliability and adaptability, ensuring safe and stable train-ground communication.

At the chip level, **integrated microresonator-based photonic microwave sensors** combined with ML and DL are becoming smaller, smarter, and capable of multiple-dimension sensing. AI models such as support vector regression (SVR), convolutional neural networks (CNNs), and neural tangent kernels (NTKs) allow these photonic sensors to automatically correct temperature drift, reduce noise, and decouple multiple environmental parameters like temperature and humidity. For example, an NTK-based model improves humidity sensing accuracy by a factor of nine [111], while CNN-based photonic sensors achieve significant accuracy enhancement in liquid concentration measurements. A recent **DL-assisted dual-parameter photonic microwave sensor** further advances this concept by using a compact silicon microring resonator, to detect both temperature and humidity simultaneously [113]. This sensor delivers 3 times higher accuracy than SVR-based techniques and remains stable even under laser drift or noisy conditions.

### 2.3.4 Photonic microwave interference cancellation

In-band full-duplex wireless transmission is an effective method to double communication capacity by allowing simultaneous transmission and reception within the same frequency band [114-115]. The primary challenge in such system is self-interference which is defined as that part of the transmitted signal leaks into the receiver and mixes with the weak received signal. To mitigate this issue, self-interference cancellation (SIC) techniques are indispensable. In electronic SIC approaches, the transmitted signal is copied and weighted in amplitude and phase, and then subtracted from the received signal, to cancel the interference. But the purely electronic systems are often constrained by limited frequency-independent bandwidth in both amplitude and in phase. In contrast, photonic SIC approaches can handle much wider frequency-independent bandwidths [92-94]. Despite this advantage, photonic SIC system still require precise and rapid tuning of variable optical attenuators (VOAs) and variable optical delay lines (VODLs), which is difficult to manage through manual control.

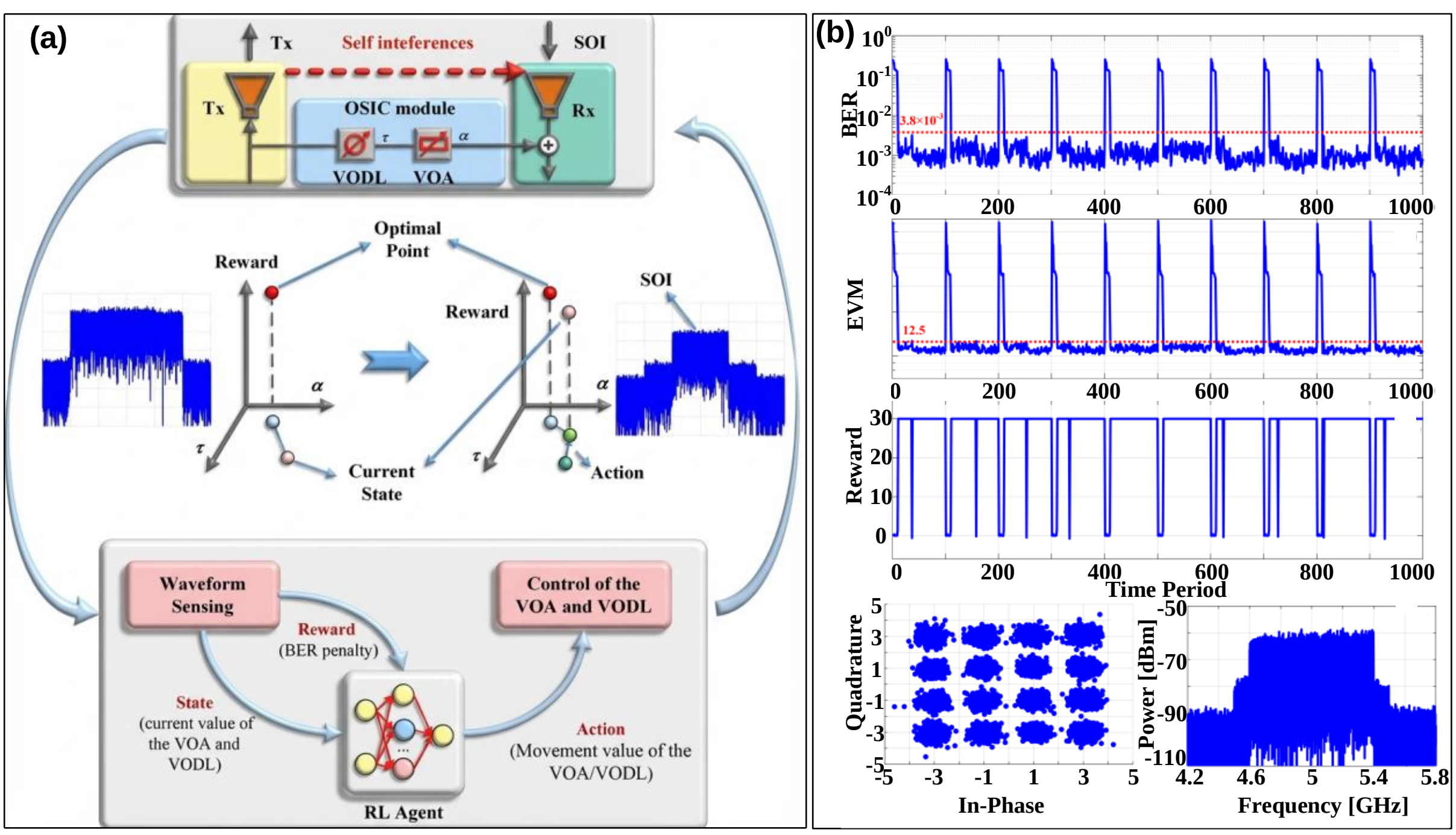

**Figure 11**. (a) Real-time adaptive optical SIC scheme using reinforcement learning algorithm and (b) measured BERs, EVMs of the recovered signal of interest [116], and corresponding rewards over time periods for a bandwidth of 800 MHz. (SIC: self-interference cancellation; VODL: variable optical delay line; VOA: variable optical attenuator).

To overcome these limitations, several **AI-enabled photonic SIC systems** have been proposed to automatically suppress the self-interference using RL. In one design as shown in **Fig. 11**, an RL agent continuously adjusts VOAs and VODLs, while the BER and the error vector magnitude (EVM) are considered as feedback [116]. This system achieves a 20.18 dB self-interference suppression over an 800 MHz bandwidth and recovers a 16-QAM OFDM signal at 5 GHz within only eight learning steps. For more complex interference environment, a multipath photonic SIC system based on deep deterministic policy gradient (DDPG) algorithm has been demonstrated [118], as shown in **Fig. 12**. It employs multiple optical paths created by wavelength-division multiplexing (WDM) and multipath optical tunable delay lines. The AI agent learns to jointly coordinate multiple optical paths to match the amplitude and the phase of the self-interference. Within five learning steps, a 24 dB self-interference suppression across a 1 GHz bandwidth is achieved and a 600 MHz 16-QAM OFDM signal is well recovered. These results demonstrate the strong adaptability and high efficiency of AI-enabled photonic SIC systems.

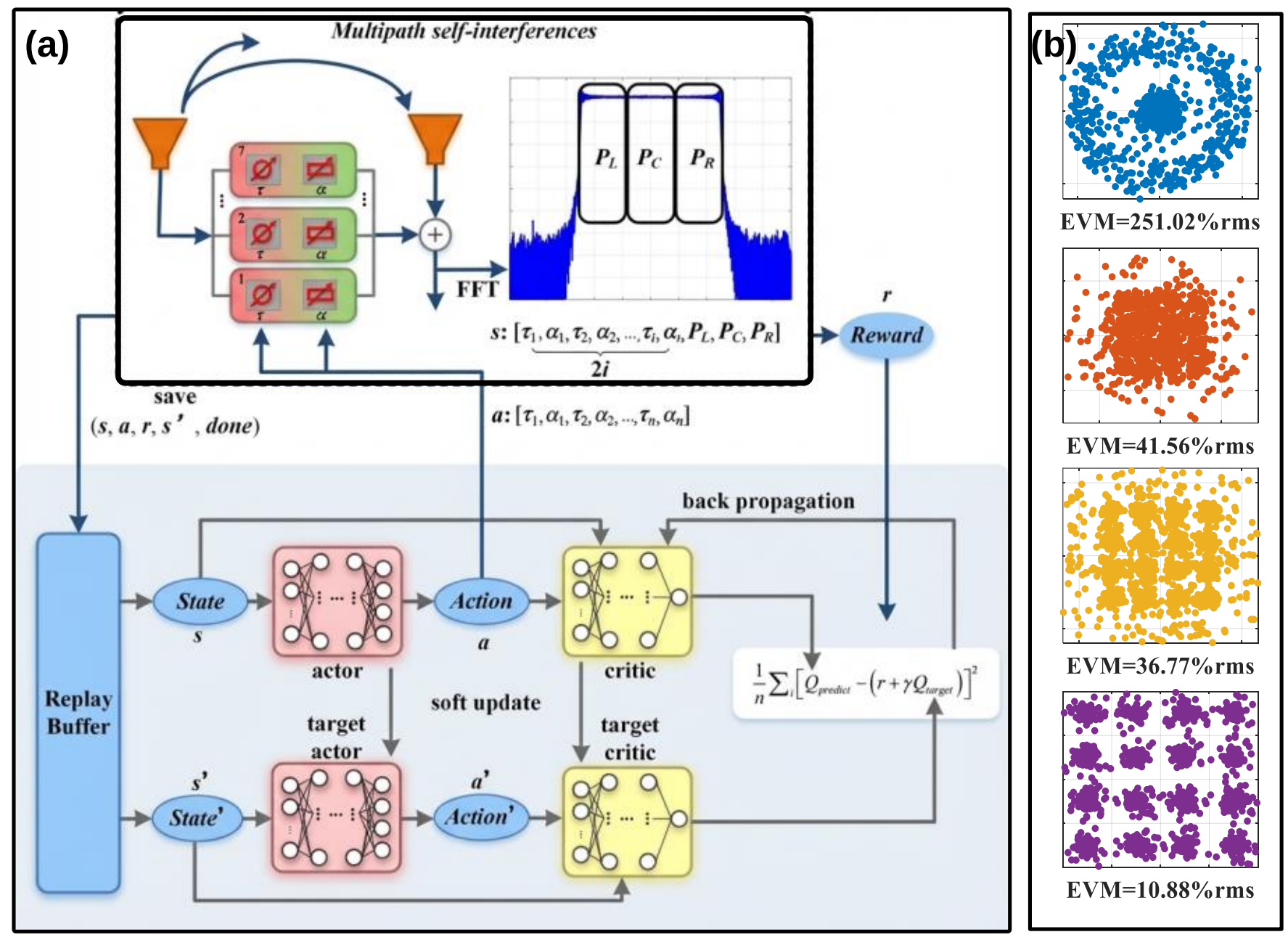

**Figure 12**. (a) Photonic-assisted wideband multipath self-interference cancellation based on DDPG strategy and (b) measured constellation diagrams of a 600-MHz 16-QAM OFDM signal without self-interference cancellation and with one-path, with two-path, and with three-path self-interference cancellation [118].

## 3. Outlook

The combination of AI and MWP marks the beginning of the **intelligent era**, referred to as **MWP 3.0**. To fully reach this goal, several key challenges still need to be addressed.

A major direction is the development of **AI co-scientist/assistant**, which can partially or fully participate in the design, testing, and optimization of MWP devices and systems. These AI agents are expected to revolutionize the MWP workflow by establishing closed-loop frameworks encompassing autonomous experiment execution, intelligent data acquisition and analysis, and adaptive system reconfiguration with minimal human intervention. By iteratively deducing optimal operational parameters for key components such as lasers, modulators, and photodetectors, these agents will enable dynamic, real-time tuning for performance optimization. When combined with digital-twin technologies, AI agents can explore and evaluate a vast number of design options in parallel, dramatically reducing the development time and cost of MWP systems.

Another promising direction is the realization of **AI-embedded MWP systems**, where intelligence is built directly into photonic hardware. Using emerging technologies such as neuromorphic photonics, on-chip optical neural networks, and intelligent metasurfaces, these systems can learn from data and experience without relying on external computation. With such smart functionalities, MWP systems will become self-calibrating, self-optimizing, and highly efficient in complex or dynamic environments.

Moreover, MWP inherently integrates multiple scientific and engineering disciplines, bridging photonics, electronics, mechanics, and thermodynamics. To achieve full-spectrum intelligence, multimodal AI agents will be required to interpret physical principles and data across these domains, as well as to integrate information from diverse modalities—including optical, microwave, millimeter-wave, acoustic, and digital signals. Such comprehensive capability will enable intelligent coordination, cross-domain learning, and global optimization across all levels of MWP systems.

## 4 Conclusion

This review has systematically summarized recent advances in **AI-enabled MWP** and outlined the evolution toward the **intelligent era (MWP 3.0)**. By deeply integrating AI with photonic technologies, MWP systems are being transformed from static, manually tuned architectures into adaptive, data-driven, and self-optimizing platforms. AI provides powerful tools for modeling, design, and optimization, enabling breakthroughs in microwave signal generation, transmission, processing, and detection with superior accuracy and efficiency. Through data-driven learning and intelligent optimization, AI empowers MWP systems to autonomously identify optimal operating conditions, compensate for complex nonlinearities, and adapt to dynamically changing environments. These capabilities significantly enhance system stability, scalability, and energy efficiency while reducing human intervention and development cost. As a result, the synergy between AI and MWP not only pushes the limits of high-frequency photonic performance but also establishes a new foundation for intelligent, reconfigurable, and fully autonomous photonic–microwave systems.